\magnification\magstep1
\font\BBig=cmr10 scaled\magstep2
\font\small=cmr7


\def\title{
{\bf\BBig
\centerline{Vortex solutions of the Liouville equation}
}
} 


\def\authors{
\centerline{
P.~A.~HORV\'ATHY and J.-C.~Y\'ERA}
\bigskip
\centerline{
Laboratoire de Math\'ematiques et de Physique Th\'eorique}
\medskip
\centerline{Universit\'e de Tours}
\medskip
\centerline{Parc de Grandmont,
F--37200 TOURS (France)
\foot{e-mail: horvathy@univ-tours.fr}
}
}

\def\runningauthors{
Horv\'athy \& Y\'era
}

\def\runningtitle{
Vortex solutions of the Liouville equation
}


\voffset = 1cm 
\baselineskip = 14pt 

\headline ={
\ifnum\pageno=1\hfill
\else\ifodd\pageno\hfil\tenit\runningtitle\hfil\tenrm\folio
\else\tenrm\folio\hfil\tenit\runningauthors\hfil
\fi
\fi}

\nopagenumbers
\footline={\hfil} 


\def\and{\qquad\hbox{and}\qquad}

\def\kikezd{\parag\underbar} 
\def\IC{{\bf C}}
\def\IR{{\bf R}}
\def\smallover#1/#2{\hbox{$\textstyle{#1\over#2}$}}
\def\2{{\smallover 1/2}}
\def\parag{\hfil\break} 
\def\={\!=\!}
\def\p{\partial}
\def\bp{\bar{\partial}}

\def\deg{ {\rm \,deg\, }}
\def\smallcirc{{\raise 0.5pt \hbox{$\scriptstyle\circ$}}}

\def\sk{({\rm sign\ }\kappa)}

\newcount\ch 
\newcount\eq 
\newcount\foo 
\newcount\ref 

\def\chapter#1{
\parag\eq = 1\advance\ch by 1{\bf\the\ch.\enskip#1}
}

\def\equation{
\leqno(\the\ch.\the\eq)\global\advance\eq by 1
}

\def\foot#1{
\footnote{($^{\the\foo}$)}{#1}\advance\foo by 1
} 

\def\reference{
\parag [\number\ref]\ \advance\ref by 1
}

\ch = 0 
\foo = 1 
\ref = 1 


\title
\vskip10mm
\authors
\vskip.25in

\parag
{\bf Abstract.}
{\it  The most general vortex solution of the Liouville equation
(which arises in non-relativistic Chern-Simons theory) is associated
with rational functions, $f(z)=P(z)/Q(z)$ where
$P(z)$ and $Q(z)$ are  both polynomials, $\deg P<\deg Q\equiv N$. 
This allows us to prove that the solution depends on 
$4N$ parameters 
without the use of an index theorem,
as well as the flux quantization~: 
$\Phi=-4N\pi\sk$}.

\vskip15mm
\noindent
(\the\day/\the\month/\the\year)
\vskip15mm
\medskip\noindent
to appear in {\sl Lett. Math. Phys}. 
\vskip15mm 

\noindent
\vskip5mm
\vfill\eject

\chapter{Vortex solutions}

Non-relativistic Chern-Simons theory supports  vortices
 [1]. (See also [2] for reviews).
For suitable values of the 
parameters, these vortices arise as solutions of the first-order 
``self-duality'' (SD) equations,
$$\big(D_{1}\pm iD_{2}\big)\psi=0,
\qquad
\kappa B=-\varrho,
\equation
$$
where $\psi$ is a complex scalar field minimally coupled to a
static gauge potential  $\vec{A}=(A_{x}, A_{x})$ in the plane; 
$D_{j}\psi=\big(\p_{j}-iA_{j}\big)\psi$ is the covariant derivative,
$B=\p_{x}A_{y}-\p_{y}A_{x}$ is the magnetic field, and
$\varrho=\psi^*\psi$ is the particle density. The solutions 
we are interested in are ``non-topological'' in that
$\varrho\to0$ at infinity.

Using the complex notations $\p=\2(\p_{x}-i\p_{y})$,
$A=A_{x}-iA_{y}$, $z=x+iy$, the first SD equation are solved by
$A=-2i\p\ln\psi$.
Setting 
$\psi=\varrho^{1/2}e^{i\omega}$, we get
$$
A=-i\p\ln\varrho+2\p\,\omega.
\equation
$$
Reinserting $A$ into the second SD equation, we end up with the
Liouville equation,
$$
\bigtriangleup\ln\varrho=-{2\over\vert\kappa\vert}\varrho,
\equation
$$ 
whose all real solutions are known. They are in fact given in terms of 
an {\it arbitrary} analytic function on the complex plane,
$$
\varrho=4\vert\kappa\vert{\vert f'(z)\vert^2\over(1+\vert 
f(z)\vert^2)^2}.
\equation
$$

For $f(z)=z^{-N}$, for example, we get the 
radially symmetric solution 
$$
\varrho=4\vert\kappa\vert N^2 
{r^{-2(N+1)}\over(1+r^{-2N})^2}.
\equation
$$
To be regular, $N$ has to be here  an integer at least $1$ [1], [2].

A striking feature of these non-topological vortex solutions
 is that their magnetix flux is an 
{\it even multiple} of the elementary flux quantum, 
$$
\Phi\equiv\int\!B\,d^2x
=2N\times\Phi_{0},
\qquad
\Phi_{0}=-2\pi,
\qquad
N=0,\pm1,\dots
\equation
$$
This can be seen easily in the radial case, 
taking into account the asymptotic behaviour, 
$\varrho\propto r^{-2(N+1)}$
as  $r\to\infty$,
of the particle density.
We are not aware of a general proof, though.

Another peculiarity of these vortices is that, for fixed $N$, the 
solution depends on $4N$ parameters. 
The proof given by 
Kim et al.  [3] uses the Atiyah-Singer index theorem.
If such an approach is perfectly justified
 for ``Nielsen-Olesen'' vortices or ``BPS'' monopoles [4],
it seems too-powerful when, as in our case, 
all solutions are known explicitly.

A $4N$-parameter 
family of solutions can be written down at once: 
consider
$$
f(z)=\sum_{i=1}^N{c_i\over z-z_{i}},
\qquad
c_i,\; z_i\in \IC,\quad
i=1,\dots,N.
\equation
$$
Plotting the  particle density allows us to interpret the associated 
solution as representing $N$ separated vortices 
located at the points $z_{i}$,  with individual scales and 
phases $c_{i}$
[5]. 
With some work, the associated flux is found to be $\Phi=-4N\pi\sk$ 
[1].

The questions we address ourselves sounds: what is the most general
$f$ yielding physically admissible solutions ? 
What is the associated magnetic flux~?
Can we count the parameters without an index theorem~?
Below, we show that 
imposing suitable regularity conditions only allows 
{\it rational functions}. In detail, we prove 

\kikezd{Theorem 1}.
 {\it Let us consider a vortex solution of the 
Liouville equation with finite magnetic flux $\Phi<\infty$.
If the magnetic field $B$ is regular on the complex plane and is such that
$
r^{2+\delta}B
$
($\delta>0$)
is bounded when $r\to\infty$, then $f$ is a rational function,
$$
f(z)={P(z)\over Q(z)},
\equation
$$
where $P(z)$ and $Q(z)$ are polynomials with $\deg P<\deg Q$. 
The  coefficient of the highest-order term  in $Q(z)$ can be normalized to $1$}.
\vskip2mm

It follows that the general solution does indeed depend on
$4N$ parameters, namely on the $2\,\times \,N$ complex coefficients of 
the polynomials with $P(z)$ and $Q(z)$.
The particular form (1.7) is recovered by expanding $f$ into 
partial fractions, provided $Q(z)$ only has simple zeroes.
 
Next we prove

\kikezd{Theorem 2}.
 {\it The magnetic flux of the vortex associated 
with the rational function
$f(z)$ in Eq. (1.8) is $\Phi=-4N\pi\sk$, where
$N$ is the degree of the denominator,
$$
N=\deg Q(z)
\equation
$$  
and is hence an integer}.

\vskip2mm
As a corollary, we also get the general theorem on flux quantization,
as in Eq. (1.6).

\goodbreak
We establish our theorems by elementary complex analysis [6], [7],
as the result of a series of Lemmas.
We start with proving the

\kikezd{Lemma 1}:
{\it Let $f(z)$ be a complex function which only has
isolated singularities. Let $\gamma$ be a curve
in the complex plane which avoids the singularities of $f$;
set $z_{0}=f(0)$ and $z_1=f(1)$. Then
$$
{\vert f(z_0)-f(z_1)\vert\over
\sqrt{1+\vert f(z_0)\vert^2}\sqrt{1+\vert f(z_1)\vert^2}}
\;\leq\;
\int_{\gamma}{2|f'(z)|\over1+|f(z)|^2}|dz|,
\equation
$$
where $|dz|=\left\vert{d\gamma/dt}dt\right\vert$}.
\vskip2mm

This proposition has a nice geometric meaning: the left-hand side
is the length of $\gamma$ with respect to a metric inherited by
stereographic projection, while the right-hand-side is the integral of
$\varrho^{1/2}$, the square-root of the particle density.

Next, using  Lemma 1, we demonstrate

\kikezd{Lemma 2}. 
{\it The function $f$ can not have an essential singularity in the complex 
plane}.
\vskip2mm

This lemma eliminates functions like
$f(z)=e^{1/z}$. 
This function would yield in fact a density which is unbounded
at the origin.
Writing  $z=r\exp{(i\theta)}$, the density becomes
$$
\varrho={4\exp{({2\cos{\theta}\over r})}\over
r^4\Big(1+\exp{({2\cos{\theta}\over r})}\Big)^2},
$$
so that on the imaginary axis, $\theta=\pi/2$, 
$\displaystyle{\lim_{r\to0}}\,\varrho=+\infty$.

\goodbreak
Next, we prove

\kikezd{Lemma 3}. {\it The function $f$ 
can not have an essential singularity at infinity}.
\vskip2mm

Lemma 3. rules out the functions  $f(z)$ as $e^z$. For this choice,
the particle density reads
$$
\varrho={4\over \Big(\exp{(-r\cos{\theta})}+
\exp{(r\cos{\theta})}\Big)^2}.
$$
On the imaginary axis $\theta=\pi/2$, we have $\varrho=4$:
the particle density is not localized.
\goodbreak

Now a theorem found in Whittaker and Watson [6], 5.64., p.105. says that
{\it the only one-valued functions which have no 
singularities, except poles, at any point (including $\infty$), are
rational functions}.
This allows us to conclude that our $f(z)$ is indeed rational. Then,

\kikezd{Lemma 4}. {\it The polynomials in Eqn. (1.8) can be chosen
so that $\deg P<\deg Q$,  the highest term in $Q$ having 
coefficient equal to $1$}.
\vskip2mm
\noindent  
This proves our Theorem 1.
\vskip2mm
\goodbreak

Now, to evaluate the magnetic flux of the vortex associated with
the rational function $f$ in Eq. (1.8), we  show that 

\goodbreak
\kikezd{Lemma 5}. {\it Let $z_{1}, \dots z_{N_{Q}}$ denote the 
distinct roots of the denominator $Q(z)$, 
each having multiplicity $n_{i}$.
The particle density (1.4) can also be written
as
$$
\varrho=4\vert\kappa\vert\bp\,\left\lbrack
\big({\partial f\over f}\big)\,
{\vert f\vert^2\over1+\vert f\vert^2}+
\sum_{i=1}^{N_{Q}}{n_{i}\over z-z_{i}}\right\rbrack,
\equation
$$
where 
the bracketed quantity is a regular function on the plane}.

\vskip2mm
Then the flux  (1.6) is converted into a contour integral at infinity,
by Stokes' theorem,
$$
\Phi=
2i\sk\oint_{S}\left\lbrack\big({\partial f\over f}\big)\,
{\vert f\vert^2\over1+\vert f\vert^2}
+\sum_{i=1}^{N_{Q}}{n_{i}\over z-z_{i}}\right\rbrack\,dz,
\equation
$$
where ${S}\equiv{S_{\infty}}$ is the cercle at infinity.

The integrand of (1.12) is  related to the vector potential.
Using (1.2), this  latter reads in fact
$$
A=2i\big({\p f\over f}\big){\vert f\vert^2\over1+\vert f\vert^2}
-i\big({\p^2f\over\p f}\big)+2\omega.
\equation
$$
To get a regular $A$, the phase $\omega$ has to be chosen so that
$$
2\p\omega=
\sum_{i=1}^{N_{Q}} {n_{i}-1\over z-z_{i}}
+
\sum_{i=1}^{N_{P}} {m_{i}-1\over z-Z_{i}},
\equation
$$
so that the integrand in (1.12) is
$$
A+\left\{i\big({\p^2f\over\p f}\big)+
i\Big(\sum_{i=1}^{N_{Q}}{n_{i}+1\over z-z_{i}}
-
\sum_{i=1}^{N_{P}}{m_{i}-1\over z-Z_{i}}
\Big)\right\};
$$
the integral of the terms in the curly bracket on the circle at infinity
vanishes.

Now using

\kikezd{Lemma 6}. 
$$
\oint_{S}\big({\partial f\over f}\big){\vert f\vert^2\over1+\vert 
f\vert^2}=0,
\equation
$$

\vskip2mm
\noindent
the  second term in (1.12) is
evaluated at once,
$$
\Phi=
2i\oint_{S}
\sum_{i=1}^{N_{Q}} {n_{i}\over z-z_{i}}\,dz
=-4\pi(\sum_{i=1}^{N_{Q}} n_i)=-4N\pi\sk,
\equation
$$
where $N$ is the degree of $Q(z)$.
This yields Theorem 2.

The  flux has been previously 
related to the inversions [8]. 
Their argument goes as follows: the particle 
density behaves as $\rho\sim r^{2(N-1)}$ when $r\to0$.
The regularity of the vector potential requires the phase to be chosen as 
$\omega=(N-1)\theta$.
Then the inversion symmetry implies the behaviour $\rho\sim r^{-2(N+1)}$
at infinity, so that the flux is indeed\quad
$
2\pi(N-1)+2\pi(N+1)=4\pi N.
$

This argument is only valid in the radial case, though. To see this, let
observe that the choice
$f(z)=(1+z)^{-2}-2(z-1)^{-1}$, yields, for example, flux
$\Phi=-12\pi$ (i.e. $N=3$) and is interpreted as  a $2$-vortex sitting
at $z=-1$ and a $1$-vortex sitting at $z=1$. The particle density 
does {\it not} behave as claimed by Kim et al [8], rather as 
 $\rho\sim r^{6}$ (instead of $r^{4}$) when $r\to0$, and 
as $\rho\sim r^{-4}$ (instead of $r^{-8}$) when $r\to\infty$.

Where does the error come from~?
On the one hand, the behaviour at the origin
assumed by Kim et al. is consistent with our formul{\ae}
 (1.4), (1.14) in the radial case only.
On the other hand, albeit 
the Liouville {\it equation} is indeed inversion-invariant
(indeed invariant with respect to any conformal transformation)
this is not true for individual, non-radial {\it solutions}.
Therefore, the large-$r$ behaviour of a solution
can not be inferred from that
for small $r$.

In this paper, we only considered the case of single-valued
functions $f$. It seems however, that multiple-valued function do not 
qualify. For example, the charge density $\varrho$ associated  with
$
f(z)=\ln z,
$
is also multiple-valued, and hence physically inadmissible;
remember also that in the radial case $f(z)=z^{-N}$, the regularity 
requires $N$ to be an integer.

\goodbreak

\vfill\eject
\chapter{Proofs.}

\kikezd{Proof of Lemma 1}.
 Stereographic projection carries 
over the natural metric from the Riemann sphere to the complex plane.
The scalar product of two tangent vectors, $u$ ad $v$, at a point $p$ 
of the plane is
$$
g_{p}(u, v)={4\over (1+\vert p\vert^2)^2}\;u\cdot v,
\equation
$$
where ``$\,\cdot\,$'' is the ordinary scalar product in $\IR^2$,
$\vert p\vert^2=p\cdot p$.
The length of a curve
 $[0,1]\ni t\mapsto\Gamma(t)\in\IC$ 
w. r. to this metric is
$$
L(\Gamma)=\int_{0}^1\!\Big\Vert{d\Gamma\over dt}\Big\Vert
dt,
\qquad\quad
\Big\Vert{d\Gamma\over dt}\Big\Vert=
\left[
g_{\Gamma(t)}
\big(\smallover{d\Gamma}/{dt},\smallover{d\Gamma}/{dt}\big)
\right]^{1/2}
=
{2\vert\Gamma'(t)\vert\over1+\vert\Gamma(t)\vert^2}.
\equation
$$
Then the distance of two points, $w_0$ et $w_1$,
in the complex plane is the
l'infimum of length of the curves between the points,
$$
d\big(w_{0}, w_{1}\big)=\hbox{\rm Inf}_{\Gamma}
\Big\{L(\Gamma)
\;\big\vert\;
\Gamma(0)=w_{0},\; \Gamma(1)=w_1
\Big\}.
$$

Let us now consider an analytic function
$
w=f(z).
$ 
$f$ can also be viewed as a mapping
of the  $z$-plane into the $w$-plane; the latter is endowed with the
distance defined here above.
If $\gamma(t)$ is an arbitrary curve in the $z$-plane 
with end-points $z_{0}$ and $z_{1}$, 
its  image by $f$ is a curve
 $\Gamma=f\smallcirc\gamma$
in the $w$-plane with end-points 
 $w_{0}=f\big(z_{0}\big)$ and
$w_{1}=f\big(z_{1}\big)$.
By (2.2), the length of
$\Gamma$ is the r. h. s. of (1.10),
$$
L(\Gamma)
=
\int_{\gamma}{2|f'(z)|\over1+|f(z)|^2}|dz|.
\equation
$$
Thus
$$
d\big(w_{0}, w_{1}\big)=
d\big(f(z_{0}), f(z_{1})\big)
\leq
\int_{\gamma}{2|f'(z)|\over1+|f(z)|^2}|dz|.
\equation
$$

But the  distance on the $w$-plane is just the distance
on the  Riemann sphere. But this latter sits in $\IR^3$,
so that the (geodesic) distance on the sphere is greater or equal to 
the natural distance in $\IR^3$:
$$
d\big(w_{0}, w_{1}\big)
\geq
{\big\vert w_{0}-w_{1}\big\vert
\over
\sqrt{1+\vert w_{0}\vert^2}\
\sqrt{1+\vert w_{1}\vert^{2}}
},
\equation
$$
equality being only achieved for  $w_{1}= w_{0}$.
Setting $w_{i}= f(z_{i}),\ i=0,1$,  the inequality (1.10)
is obtained.

\hfill Q.~E.~D.
\goodbreak

\kikezd{Proof of Lemma 2}.
Let us assume that $f$ has an isolated essential singularity at a 
point $z_0$. Then it is analytic in some disk
$D\equiv D(z_0;\epsilon)\setminus\lbrace z_0\rbrace$. 
Now, according to 
Picard's Theorem ([7], p. 90): 
{\it If $z_0$ is an isolated singularity of a 
holomorphic function $f(z)$, then for each $r>0$, l'image of the
annular region $\big\{z\in \IC \big\vert 0<\vert z-z_0\vert< 
r\big\}$ is either the whole of $\IC$ or $\IC$ without a single point}.

\vskip1mm
Let us first assume that
 $z_1$ is a point in $D$ such that $f(z_1)=0$.
Then, since the particle density,
$\varrho$,  is a regular  function on the plane which goes to zero at
infinity, there is a real number $M$ such that
$$
\rho(z)={4|f'(z)|^2\over(1+|f(z)|^2)^2}\leq M^2,
\qquad\forall z\in \IC.
$$

Eqn. (1.10) in Lemma 1. with $z_{0}=z$, yields, using  $f(z_{1})=0$,
$$
{|f(z)|\over\sqrt{1+|f(z)|^2}}\leq M\int_{\gamma}|dz|
$$
for all curve $\gamma$ s. t. $\gamma(0)=z$ et $\gamma(1)=z_1
$.
For the straight line
$\gamma(t)=z+t(z_1-z)$ in particular, the r. h. s.
becomes
$\vert z_{1}-z\vert M<2\epsilon M$, since $z_{1}$ and $z$ both belong 
to $D$. 
Thus, chosing 
 $\epsilon$ to have $4M\epsilon\leq1$,
$
{|f(z)|/\sqrt{1+|f(z)|^2}}<{1/2},
$ 
which implies that
$
|f(z)|\leq 1.
$
The function $f(z)$ is hence bounded in $D$,
which contradicts hypothesis that $z_0$ is an essential singularity.

Now if $f$ does not vanish in
 $D\equiv D(z_0;\epsilon)$,
one can chose $z_0'$ with $f(z_0')$ sufficiently small so that
$f(z)$ is still bounded in $D$.

\vskip1mm
\hfill Q.~E.~D.
\goodbreak

\kikezd{Proof of Lemma 3}. Let us assume, on the contrary, that
infinity is  essential singularity of $f(z)$. 
From the large-$r$ behaviour $B=o(r^{2})$ we shall deduce that 
$f$ is again bounded at infinity, a contradiction.

In fact, if $\infty$ is an isolated singularity, then 
$f(z)$ is holomorphic in some neighbourhood 
 $D\equiv\lbrace z\in\IC\,\Big\vert\, \vert z\vert>N\rbrace$, 
of infinity.
In this neighbourhood, one can find a point $z_0$ where $f$ vanishes 
$f(z_0)=0$ by Picard's theorem
\foot{If $f(z)$ never vanishes in the region $D$,
it is enough to chose $f(z_{0})$ small enough.}.

Now, due to the imposed groth condition on $B=-\varrho/\kappa$,
$N$ can be taken so that
$$
\varrho\leq C^2\vert z\vert^{-2-\delta},
\qquad\hbox{for all}\quad
\vert z\vert > N,
\equation
$$
where $C>0$ is a  constant.
Let us chose
 $N$ such that $\displaystyle{ 4C(\pi+{1\over\delta})<N^{\delta/2}}$.
Then, for all complex number $z$, $\vert z\vert>\vert z_0\vert$,
$f$ is bounded,
$
\vert f(z)\vert\leq1.
$
To see this, consider $z_1$,  the intersection of $C$, the cercle
around $0$ with radius  $\vert z\vert$, with the
straight half-line $\lbrack0,z_0)$. Then
$$
d(0,f(z))\leq d(0,f(z_1))+d(f(z_1),f(z)).
\equation
$$
Now, applying the inequality  (2.4) to the circular arc
 $(z_1,z)$ we get
$$
d(f(z_1),f(z))\leq \int_{\small arc}\varrho(f(z))^{1/2}\vert dz\vert.
\equation
$$
From this we deduce, using (2.6), that
$$
d(f(z_1),f(z))\leq{2\pi C\over N^{\delta/2}}.
\equation
$$ 
On the other hand, applying (2.4) to the segment
$\gamma(t)=z_0+t(z_{1}-z_0),\ t\in\lbrack0,1\rbrack$, we get
$$
d(0,f(z_1))\leq\int_{\gamma}\varrho\big(f(z)\big)^{1/2}
\vert dz\vert
\leq\int_{\gamma}{C\over \vert z\vert^{1+\delta/2}}\vert dz\vert
\leq{ 2C\over\delta N^{\delta/2}},
\equation
$$
when the condition (2.6) is used again.
The  inequalities (2.7)--(2.9)--(2.10) imply that
$$
d(0,f(z))\leq {2C\over N^{\delta/2}} 
(\pi+{1\over\delta})<{1\over2};
\equation
$$
Now, by (2.5), we have
$$
d(0, f(z))
\geq
{\vert f(z)\vert\over\sqrt{1+\vert f(z)\vert^2}};
$$
using (2.11), we get finally $f(z)\vert\leq1$.
The function $f(z)$ is hence bounded in some neighbourhood of
infinity, so that this point cannot be an essential singularity.
\vskip1mm
\hfill Q.~E.~D.
\goodbreak

Now (as explained in Chapter 1), Theorem 5.64 of
Whittaker and Watson [6] allows us to deduce that $f$ is a rational 
function, $f(z)=P(z)/Q(z)$, where $P(z)$ and $Q(z)$ are both 
polynomials.

\kikezd{Proof of Lemma 4}. 
Now, since $f$ and $f^{-1}$ are readily seen to yield the same solutions,
we can assume that $\deg P\leq\deg Q$. The case $\deg P=\deg Q$
is eliminated by a simple redefinition,  as in Ref. [1]. 
In fact, 
$$
f(z)=f_0+{A(z)\over B(z)},
$$
where $A(z)$ are $B(z)$ polynomials s. t. $\deg{A(z)}<\deg{B(z)}$
and $f_0\neq0$, is readily seen
to yield the same density (1.4) as
$$
\tilde{f}(z)={\tilde{A}(z)\over\tilde{B}(z)},
$$
with the polynomials
$\tilde{A}(z)$ and $\tilde{B}(z)$ defined as
$$
\tilde{A}(z)={A(z)\over1+\vert f_0\vert^2},
\qquad
\tilde{B}(z)=B(z)+\bigg({\bar{f_0}\over1+\vert f_0\vert^2}\bigg)A(z).
$$ 
Then the coefficient of the highest-order term in $Q(z)$ can be 
normalized to unity.

\vskip1mm
\hfill Q.~E.~D.
\goodbreak

\kikezd{Proof of Lemma 5}. 
In complex notations, 
the general solution (1.4) is expressed as
$$
\varrho=4\vert\kappa\vert\bar{\partial}\,\left\lbrack
\big({\partial f\over f}\big)\,
{\vert f\vert^2\over1+\vert f\vert^2}\right\rbrack,
\equation
$$
where $\bp=\2\big(\p_{x}+i\p_{y}\big)$. 
Let us now denote by $z_{i},\, i=1,\dots, N_{Q}$ the distinct roots of the 
denominator, $Q(z)$, each having a multiplicity $n_i$.
[Since $P(z)$ and $Q(z)$ have no commun roots, these are the same as 
the the poles of $f(z)$]. Then the function
$$
\big({\partial f\over f}\big)
{\vert f\vert^2\over1+\vert f\vert^2}+
\sum_{i=1}^{N_{Q}} {n_{i}\over z-z_{i}}
\equation
$$
is regular on the complex plane.
Indeed, in the neighbourhood of a root, $z_{i}$, of order 
$n_{i}$ of $Q(z)$,
$$
f\sim{c_{i}\over(z-z_{i})^{n_i}}
\qquad\Longrightarrow\qquad
{\p f\over f}\sim-{n_{i}\over(z-z_{i})},
\quad
{\vert f\vert^2\over1+\vert f\vert^2}\sim1.
$$

In contradistinction,  in the
neighbourhood of a zero (denoted by $Z_{0}$)
of order $k\geq1$, of $P(z)$ [which is the same as a 
zero of $f(z)$], we have~:
$$
f\sim(z-Z_{0})^k
\qquad\Longrightarrow\qquad
{\p f\over f}\sim {k\over z-Z_{0}},
\quad
{\vert f\vert^2\over1+\vert f\vert^2}\sim \vert z-Z_{0}\vert^{2k},
$$
so that (2.13) is  regular.
Now, since
$$
\bp\left(\sum_{i=1}^{N_{p}} {n_{i}\over z-z_{i}}\right)=0,
$$
the second term in (2.13) can be added to the expression (2.12)
of the density, yielding (1.11).

\vskip1mm
\hfill Q.~E.~D.
\goodbreak

The integrand in (2.13) being a regular function, the flux 
$\Phi=-(1/\kappa)\int\!\varrho\,d^2x$
can be converted, by Stokes theorem, into an integral at infinity,
to yield (1.12).

\kikezd{Proof of Lemma 6}. Let us now denote by $z_{i}$, 
$i=1,\dots,N_{P}$, the distinct roots of the numerator, $P(z)$
[and hence those of $f(z)$].
 Using the theorem on the residues,
$$
\oint_{S}\big({\partial f\over f}){\vert f\vert^2\over1+\vert f\vert^2}dz=
\left(\lim_{\vert z\vert\to\infty}{\vert f(z)\vert^2\over1+\vert 
f(z)\vert^2}\right)\,2\pi i\,
\left(\sum_{i=1}^{N_{P}} m_{i}-\sum_{i=1}^{N_{Q}} n_{i}\right)=0,
$$
since $\displaystyle{\lim_{\vert z\vert\to+\infty}}f(z)=0$.
\vskip1mm
\hfill Q.~E.~D.
\goodbreak

\kikezd{Acknowledgements}.
J.-C. Y. acknowledges
the {\it Laboratoire de Math\'emathiques et de Physi\-que Th\'eori\-que}
of Tours University for hospitality.  
He is also indebted to the {\it Gouvernement de La C\^ote d'Ivoire}
for a doctoral scholarship.
The authors are indebted to Professor R. Jackiw 
for illuminating discussions.

\vskip5mm
\goodbreak


\centerline{\bf\BBig References}

\reference 
R.~Jackiw and S-Y.~Pi,
{\sl Phys. Rev. Lett}. {\bf 64}, 2969 (1990);
{\sl Phys. Rev}. {\bf D42}, 3500 (1990).

\reference 
R. Jackiw and S-Y. Pi, 
{\sl Prog. Theor. Phys. Suppl}. {\bf 107}, 1 (1992),
or
G. Dunne, {\sl Self-Dual Chern-Simons Theories}. 
Springer Lecture Notes in Physics. New Series: Monograph 36. (1995).

\reference 
S. K. Kim, K. S. Soh, and J. H. Yee,
{\sl Phys. Rev}. {\bf D42}, 4139 (1990).

\reference 
E. Weinberg, 
{\sl Phys. Rev.} {\bf D 19}, 3008
(1979).

\reference 
P.~Horv\'athy, J.~C. Y\'era, 
{\sl Phys. Rev.} {\bf D 54}, 4171 (1996).

\reference 
E.~T.~Whittaker and G. N. Watson,
{\sl A course of modern analysis}.
Fourth edition. Cambridge University Press.
Reprinted in 1992.

\reference 
H. Cartan, {\it Th\'eorie \'el\'ementaire des fonctions analytiques 
d'une ou plusieurs variables complexes}, Sixi\`eme Edition,
Hermann (1985).

\reference 
S. K. Kim, W. Namgung, K. S. Soh, and J. H. Yee,
{\sl Phys. Rev}. {\bf D46}, 1882 (1992).

\bye